\documentclass[aps,prb,twocolumn,noshowkeys]{revtex4}

\usepackage{graphicx,color}
\usepackage{bm}
\usepackage{dcolumn}
\usepackage{amsmath}

\begin{document}
\title{Coherence lengths for superconductivity in the two-orbital negative-$U$ Hubbard model}

\author{Grzegorz Litak}
\affiliation{
Lublin University of Technology, Faculty of Mechanical Engineering, Nadbystrzycka 36,
PL-20618 Lublin, Poland
}

\author{Teet \"Ord}
\affiliation{
Institute of Physics, University of Tartu, T\"{a}he 4, 51010 Tartu, Estonia
}

\author{K\"ullike R\"ago}
\affiliation{
Institute of Physics, University of Tartu, T\"{a}he 4, 51010 Tartu, Estonia
}

\author{Artjom Vargunin}
\affiliation{
Institute of Physics, University of Tartu, T\"{a}he 4, 51010 Tartu, Estonia
}



\begin{abstract}

We study the peculiarities of coherency in the superconductivity of two-orbital system. The superconducting phase transition is caused here by the on-site intra-orbital
attractions (negative-$U$ Hubbard model) and inter-orbital pair-transfer interaction. The dependencies of critical and non-critical correlation lengths on interaction
channels and band fillings are analyzed. 
\end{abstract}

\maketitle

\section{Introduction}

Multi-component superconductivity models, developed more than fifty years starting from the papers [\onlinecite{smw, m, kondo}], include rather varied physics.
In connection with the presence of interacting order parameters the
main properties of the multi-band systems are quite different from the corresponding characteristics in single-band superconductors. The examination of theses peculiarities has been an object of growing interest.

The various multi-component theoretical scenarios have been applied for a number of superconducting materials
(see Refs. [\onlinecite{kw, kko, bianconi1, bianconi2}] and references therein). The research activity in this direction has been especially stimulated by the
 acceptance of the multi-gap superconductivity in MgB$_2$ \cite{MgB2}, cuprates \cite{cuprates} and iron-arsenic compounds \cite{pnictides}. In
 particular, the derivation of high-quality superconducting regions from oxygen ordering, observed recently in La$_{2}$CuO$_{4+y}$
\cite{oxygen}, supports the multi-band theoretical scenario of superconductivity in cuprates.

In the present contribution we study the coherency of the superconducting ordering
of a two-band (two-orbital) system with the negative-$U$ Hubbard intra-orbital pairing and
inter-orbital pair-transfer interaction. One can distinguish here two characteristic
length scales \cite{ord1} in the spatial behaviour of superconducting fluctuations. One of these lengths as a function of temperature behaves critically
diverging at the phase transition point. The other one remains finite and its temperature dependence is weaker. The formation of these length scales is caused
by the interband interaction mixing the superconducting order parameters of initially non-interacting bands. As a result the critical and non-critical
coherence lengths associate with critical and non-critical fluctuations (see e.g. Refs. [\onlinecite{ord2, ord3}]) which appear as the certain linear
combinations of the deviations from the equilibrium band
superconducting orders.
Consequently, these length
scales cannot be attributed to different bands involved \cite{babaev} (see also Refs. [\onlinecite{poluektov, konsin, kristoffel}]). Our results have been obtained using the  superconducting negative-$U$ Hubbard model\cite{Ref mrr} for a two-orbital system on a two-dimensional lattice.

\section{Two-orbital model of superconductivity}

We start with the Hamiltonian of the two-orbital model\cite{Ref cw} of the following form:
\begin{eqnarray}\label{1}
H&=&\sum_{\alpha}\sum_{i,j}\sum_{\sigma}\left[t_{ij}^{\alpha\alpha}+\left(\varepsilon_{\alpha}^{0}-\mu\right)
\delta_{ij}\right]a_{i\alpha\sigma}^{+}a_{j\alpha\sigma} \nonumber \\
&+&\frac{1}{2}\sum_{\alpha}\sum_{i}\sum_{\sigma}U^{\alpha\alpha} n_{i\alpha\sigma}n_{i\alpha-\sigma}\nonumber \\
&+&\frac{1}{2}\sum_{\alpha,\alpha'}^{,}\sum_{i}\sum_{\sigma}U^{\alpha\alpha'}a_{i\alpha\sigma}^{+}a_{i\alpha'\sigma}
a_{i\alpha-\sigma}^{+}a_{i\alpha'-\sigma} \, ,
\end{eqnarray}
 where $a_{i\alpha\sigma}^{+}$ ($a_{i\alpha\sigma}$) is the electron creation (destruction) operator in the orbital $\alpha=1,2$ localized at the
 site $i$; $\sigma$ is the spin index; $t_{ij}^{\alpha\alpha}$ is the hopping integral; $\varepsilon_{\alpha}^{0}$ is the orbital energy; $\mu$
 is the chemical potential; $U^{\alpha\alpha}<0$ is the intra-orbital attraction energy;
$n_{i\alpha\sigma}=a_{i\alpha\sigma}^{+}a_{i\alpha\sigma}$ is the particle number operator; $U^{\alpha\alpha'}$ with $\alpha\neq\alpha'$ is the inter-orbital
interaction energy.

The transformation of the Hamiltonian (Eq. \ref{1}) into the reciprocal space leaves us with the expression
\begin{eqnarray}\label{2}
H&=&\sum_{\alpha}\sum_{\mathbf{k}}\sum_{\sigma}\left[\varepsilon_{\alpha}(\mathbf{k})-\mu
\right]a_{\alpha\mathbf{k}\sigma}^{+}a_{\alpha\mathbf{k}\sigma} \nonumber  \\
&+&\frac{1}{2N}\sum_{\alpha,\alpha'}\sum_{\mathbf{k},\mathbf{k}'}\sum_{\mathbf{q}}\sum_{\sigma}U^{\alpha\alpha'}
a_{\alpha(\mathbf{k}+\mathbf{q})\sigma}^{+}
a_{\alpha'\mathbf{k}\sigma} \nonumber \\
&\times& a_{\alpha(\mathbf{k}'-\mathbf{q})-\sigma}^{+}a_{\alpha'\mathbf{k}'-\sigma}  \, ,
\end{eqnarray}
where $\varepsilon_{\alpha}(\mathbf{k})$ is the electron band energy 
with a wave vector $\mathbf{k}$, associated with the orbital $\alpha$ and $N$ is the
number
of lattice
sites
(number
of atoms).

For the description of spatially homogeneous superconductivity one introduces the equilibrium mean-field Hamiltonian (the terms which do not contain operators have been omitted)
\begin{eqnarray}\label{3}
H_{mf}&=&\sum_{\alpha}\sum_{\mathbf{k}}\sum_{\sigma}\tilde{\varepsilon}_{\alpha}(\mathbf{k})
a_{\alpha\mathbf{k}\sigma}^{+}a_{\alpha\mathbf{k}\sigma} \nonumber \\
&+&\frac{1}{2}\sum_{\alpha}\sum_{\mathbf{k}}\sum_{\sigma}\left(\Delta_{\alpha}a_{\alpha\mathbf{k}\sigma}^{+}
a_{\alpha-\mathbf{k}-\sigma}^{+} + h.c. \right)
\end{eqnarray}
with
\begin{equation}\label{4}
\tilde{\varepsilon}_{\alpha}(\mathbf{k})=\varepsilon_{\alpha}(\mathbf{k})+\frac{1}{2}U^{\alpha\alpha}n_{\alpha}-\mu \, .
\end{equation}
Here the average number of electrons per site
\begin{equation} \label{5}
\sum_{\alpha}n_{\alpha}=N^{-1}\sum_{\alpha}\sum_{\mathbf{k}}\sum_{\sigma}\left\langle a_{\alpha\mathbf{k}\sigma}^{+}a_{\alpha\mathbf{k}\sigma}\right\rangle_{H_{mf}}
\end{equation}
determines the position of the chemical potential $\mu$  and $\langle ... \rangle_H = Z^{-1} {\rm Sp} ... \exp{(-H/k_B T)}$
denotes
averaging procedure,
where $Z$ is a partition function, $k_B$ is the Boltzmann constant, and $T$ is temperature.
The homogeneous equilibrium superconductivity gaps are defined as
\begin{equation} \label{6}
\Delta_{\alpha}=N^{-1}\sum_{\alpha'}U^{\alpha\alpha'}\sum_{\mathbf{k}}\left\langle a_{\alpha'-\mathbf{k}\downarrow} a_{\alpha' \mathbf{k}\uparrow}\right\rangle_{H_{mf}} \, .
\end{equation}
The Eqs. (\ref{5}) and (\ref{6}) should be solved self-consistently. It is easy to find that
\begin{equation} \label{7}
\left\langle a_{\alpha\mathbf{k}\sigma}^{+}a_{\alpha\mathbf{k}\sigma}\right\rangle_{H_{mf}}=\frac{1}{2}\left[1
-\frac{\tilde{\varepsilon}_{\alpha}(\mathbf{k})}{E_{\alpha}(\mathbf{k})}\tanh\frac{E_{\alpha}(\mathbf{k})}{2k_{B}T}\right]
\end{equation}
and
\begin{equation} \label{8}
\left\langle a_{\alpha-\mathbf{k}\downarrow}a_{\alpha\mathbf{k}\uparrow}\right\rangle_{H_{mf}}=\frac{-\Delta_{\alpha}}{2E_{\alpha}(\mathbf{k})}
\tanh\frac{E_{\alpha}(\mathbf{k})}{2k_{B}T} \, ,
\end{equation}
where
\begin{equation} \label{9}
E_{\alpha}(\mathbf{k})=\sqrt{\tilde{\varepsilon}^{2}_{\alpha}(\mathbf{k})+\left|\Delta_{\alpha}\right|^{2}} \, ,
\end{equation}
The system of gap equations (\ref{6}) may be now presented in the form
\begin{equation} \label{9a}
\Delta_{\alpha}=-\sum_{\alpha'}U^{\alpha\alpha'}\Delta_{\alpha'}\zeta_{\alpha'}\left(T,\Delta_{\alpha'}\right) \,
\end{equation}
with
\begin{equation} \label{9b}
\zeta_{\alpha}\left(T,\Delta_{\alpha}\right)
=(2N)^{-1}\sum_{\mathbf{k}}\frac{1}{E_{\alpha}(\mathbf{k})}
\tanh\frac{E_{\alpha}(\mathbf{k})}{2k_{B}T} \, ,
\end{equation}
The critical temperature of the phase transition $T_c$ in the two-gap superconductor (Eq. 1) under consideration is determined by the following condition
\begin{equation}\label{9c}
\left|%
\begin{array}{cc}
  1+U^{11}g_{1}\left(T_{c}\right) & U^{12}g_{2}\left(T_{c}\right) \\
  U^{21}g_{1}\left(T_{c}\right) &  1+U^{22}g_{2}\left(T_{c}\right)\\
\end{array}%
\right|\;
=0
\end{equation}
with $g_{\alpha}\left(T_{c}\right)=\zeta_{\alpha}\left(T_{c},0\right)$.

The calculated [\onlinecite{spectrum}]  dependency of superconducting phase transition temperature  $T_c$ on band filling  $n$ is depicted in Fig. 1.
As expected, Van Hove singularity present in the middle of the band\cite{Ref markiewicz,Ref cw} is reflected as a peak in $T_c$  for half filled system
($n=1$)
where
the chemical potential $\mu$ passes this singular point.  The dependence of the small interband interaction $U^{12}$ is presented
by three separate lines   $|U^{12}|=|U^{21}|=
0.01$, 0.04, and 0.07 $t$ for '1'--'3', respectively. The larger $|U^{12}|$ increases the critical temperature $T_c$.

\begin{figure}[!h]
\includegraphics[width=40mm,angle=-90]{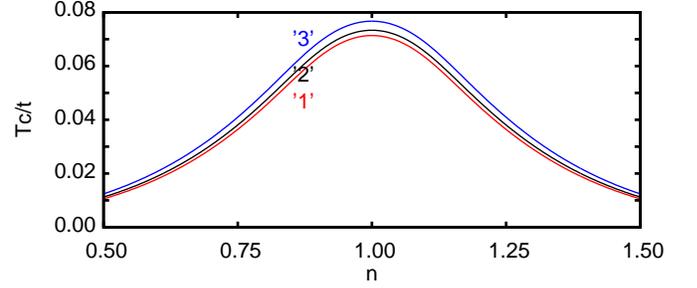}
 \caption{The critical temperature $T_c$ versus band filling 
 $n$ (equal for each band $n=n_1=n_2$) for the chosen set of interactions: $U^{11}=-1.2 t$,
$U^{22}=-1.3 t$.
$|U^{12}|=|U^{21}|$ was
0.01, 0.04, and 0.07 $t$ for '1'--'3', respectively.}
\end{figure}

\begin{figure}[!h]
\vspace{-1.0cm}
\includegraphics[width=62mm,angle=-90]{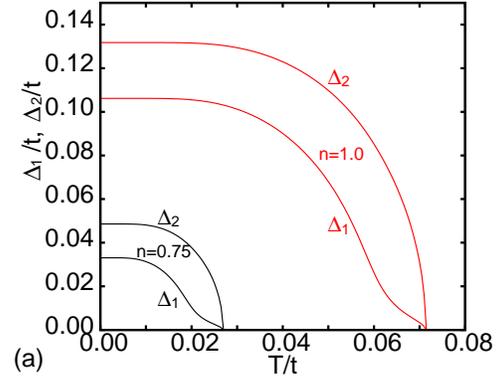}  \\
\includegraphics[width=62mm,angle=-90]{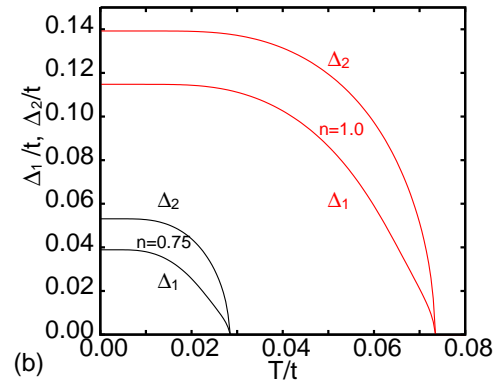}
 \caption{Superconducting gap parameters $\Delta_1$ and  $\Delta_2$ versus temperature  $T$ for various band fillings. $U^{11}=-1.2 t$, $U^{22}=-1.3 t$.
 $|U^{12}|=|U^{21}|$=0.01, 0.04 $t$, for (a) and (b), respectively. }
\end{figure}

The superconducting gaps as temperature functions are shown in Fig. 2 for various inter-orbital interactions and band fillings.
Note that for fairly low temperature ($T << T_c$), $\Delta_1$ and $\Delta_2$ are simultaneously scaled by different
intra-orbital
interactions $U_{ii}$ ($U_{11} \neq U_{22}$) and the density of states around the chemical potential $\mu$ (dependent on band filling
which was here $n=1.0$ or
0.75).  On the other hand, the influence of $U^{12}$ is visible in the vicinity of $T_c$. The stronger $U^{12}$, the larger $\Delta_1$.
This is a consequence of the "inter-orbital proximity effect".

\section{Ginzburg-Landau equations}

In the non-homogeneous situation one has to retain the non-zero momentum of Cooper pairs in the Hamiltonian (Eq. \ref{2}) which can be rewritten as
\begin{eqnarray}\label{10}
H&=&\sum_{\alpha}\sum_{\mathbf{k}}\sum_{\sigma}\left[\varepsilon_{\alpha}(\mathbf{k})-\mu
\right]a_{\alpha\mathbf{k}\sigma}^{+}a_{\alpha\mathbf{k}\sigma} \nonumber \\
&+&\frac{1}{2N}\sum_{\alpha,\alpha'}\sum_{\mathbf{k},\mathbf{k}'}\sum_{\mathbf{q}}\sum_{\sigma}U^{\alpha\alpha'}
a_{\alpha\mathbf{k}\sigma}^{+}
a_{\alpha(-\mathbf{k}+\mathbf{q})-\sigma}^{+}  \nonumber \\
&\times&a_{\alpha'(-\mathbf{k}'+\mathbf{q})-\sigma}
a_{\alpha'\mathbf{k}'\sigma} \, .
\end{eqnarray}
The Bogoliubov theorem determines the upper limit for the free energy of the system
\begin{equation} \label{11}
F_{B}=-k_{B}T\ln(Z'_{mf})+\left\langle H-H'_{mf}\right\rangle_{H'_{mf}} \, ,
\end{equation}
\begin{equation} \label{12}
Z'_{mf}=\mathrm{Sp}\exp\left(-\frac{H'_{mf}}{k_{B}T}\right) \,
\end{equation}
by means of the non-equilibrium mean-field Hamiltonian
\begin{eqnarray}\label{13}
&&H'_{mf}=\sum_{\alpha}\sum_{\mathbf{k}}\sum_{\sigma}\tilde{\varepsilon}_{\alpha}(\mathbf{k})
a_{\alpha\mathbf{k}\sigma}^{+}a_{\alpha\mathbf{k}\sigma} \nonumber \\
&&+\frac{1}{2}\sum_{\alpha}\sum_{\mathbf{k},\mathbf{q}}\sum_{\sigma}
\left(\delta_{\alpha\mathbf{q}}a_{\alpha\mathbf{k}\sigma}^{+}
a_{\alpha(-\mathbf{k}+\mathbf{q})-\sigma}^{+} + h.c. \right)
\end{eqnarray}
with non-equilibrium superconductivity gap order parameters $\delta_{\alpha\mathbf{q}}$. The free energy  upper limit $F_{B}$ can be found as an expansion in powers of $\delta_{\alpha\mathbf{q}}$ and $\mathbf{q}$. After the minimization of $F_{B}$ and certain transformations one can obtain the system of equations for the non-homogeneous superconductivity gaps $\Delta_{\alpha}(\mathbf{r})$ (the Ginzburg-Landau equations) in the following form:
\begin{eqnarray}\label{14}
\Delta_{\alpha }(\mathbf{r})&=&-\sum_{\alpha '}U^{\alpha\alpha'}\biggl[g_{\alpha'}(T)-\nu_{\alpha'}(T)
\left|\Delta_{\alpha '}(\mathbf{r})\right|^{2} \nonumber \\
&+&\sum_{l,l'=1}^{d}\beta_{\alpha 'll'}(T)\frac{\partial}{\partial x_{l}}\frac{\partial}{\partial x_{l'}}\biggr]\Delta_{\alpha '}(\mathbf{r}) \, ,
\end{eqnarray}
where $l$ refers to the Cartesian axis and $d$ is the dimension of lattice. In Eqs. (\ref{14})
\begin{eqnarray}\label{15}
g_{\alpha}(T)= \frac{1}{2N}\sum_{\mathbf{k}}\frac{1}{\tilde{\varepsilon}_{\alpha}(\mathbf{k})}
\tanh\frac{\tilde{\varepsilon}_{\alpha}(\mathbf{k})}{2k_{B}T}  \, ,
\end{eqnarray}
\begin{eqnarray}\label{16}
\nu_{\alpha}(T)= \frac{-1}{2N}\sum_{\mathbf{k}}\frac{\partial}{\partial \left|\Delta_{\alpha}\right|^{2}}\left[\frac{1}{E_{\alpha}(\mathbf{k})}
\tanh\frac{E_{\alpha}(\mathbf{k})}{2k_{B}T}\right]_{\Delta_{\alpha}=0}  \, ,
\end{eqnarray}
%

\begin{eqnarray}\label{17}
&& \beta_{\alpha ll'}(T)= \frac{-1}{4N}\sum_{\mathbf{k}}
\frac{\partial}{\partial q_{l}}\frac{\partial}{\partial q_{l'}}\left\{
\frac{1}{\tilde{\varepsilon}_{\alpha}(\mathbf{k})+\tilde{\varepsilon}_{\alpha}(\mathbf{k}-\mathbf{q})}\right . \nonumber \\
& & \left. \times  \left[\tanh\left(\frac{\tilde{\varepsilon}_{\alpha}(\mathbf{k})}{2k_{B}T}\right)
 +\tanh\left(\frac{\tilde{\varepsilon}_{\alpha}(\mathbf{k}-\mathbf{q})}{2k_{B}T}\right)\right]\right\}_{\mathbf{q}=0} \! \! \!.
\end{eqnarray}
By supposing further that
\begin{eqnarray}\label{18}
\beta_{\alpha ll'}=\beta_{\alpha}\delta_{ll'}
\end{eqnarray}
one obtains
\begin{eqnarray}\label{19}
\Delta_{\alpha }(\mathbf{r})&=&-\sum_{\alpha '}U^{\alpha\alpha'}\biggl[g_{\alpha'}(T)-\nu_{\alpha'}(T)
\left|\Delta_{\alpha '}(\mathbf{r})\right|^{2} \nonumber \\
&+&\beta_{\alpha '}(T)\sum_{l=1}^{d}\frac{\partial^{2}}{\partial x_{l}^{2}}\biggr]\Delta_{\alpha '}(\mathbf{r}) \, .
\end{eqnarray}

\section{Coherence lengths}

The small deviations from the bulk values of superconductivity gaps
\begin{eqnarray}\label{20}
\eta_{\alpha}(\mathbf{r})=\Delta_{\alpha}(\mathbf{r})-\Delta_{\alpha} \,
\end{eqnarray}
satisfy linearized equations
\begin{eqnarray}\label{21}
 \eta_{\alpha }(\mathbf{r})=-\sum_{\alpha '}U^{\alpha\alpha'}\biggl[\tilde{g}_{\alpha '}(T) +\beta_{\alpha '} (T)\sum_{l=1}^{d}\frac{\partial^{2}}{\partial
x_{l}^{2}}\biggr]\eta_{\alpha '}(\mathbf{r}),  && \nonumber \\
 &&
\end{eqnarray}
where
\begin{eqnarray}\label{22}
\tilde{g}_{\alpha }(T)=g_{\alpha}(T)-3\nu_{\alpha} (T)
\left(\Delta_{\alpha }(T)\right)^{2} \,
\end{eqnarray}
with $\Delta_{\alpha }(T)$ being the solutions of the system of equations (Eq. \ref{6}). We introduce the coherence lengths $\xi$ as the spatial length scales
of $\eta_{\alpha }(\mathbf{r})$:
\begin{eqnarray}\label{23}
\eta_{1,2}(\mathbf{r})\sim\mathrm{exp}\biggl(-\frac{\sum_{l=1}^{d}x_{l}}{\sqrt{d}\xi}\biggr) \, .
\end{eqnarray}
From Eqs. (\ref{19}) and (\ref{23}) follows the system
\begin{eqnarray}\label{24}
\biggl(1+U^{11}\tilde{g}_{1}(T)+U^{11}\beta_{1}(T)\xi^{-2}\biggr)\eta_{1}(\mathbf{r}) \nonumber \\
+\biggl(U^{12}\tilde{g}_{2}(T)
+U^{12}\beta_{2}(T)\xi^{-2}\biggr)\eta_{2}(\mathbf{r})&=&0 \nonumber\\
\biggl(U^{21}\tilde{g}_{1}(T)
+U^{21}\beta_{1}(T)\xi^{-2}\biggr)\eta_{1}(\mathbf{r}) \nonumber \\
+
\biggl(1+U^{22}\tilde{g}_{2}(T)
+U^{22}\beta_{2}(T)\xi^{-2}\biggr)\eta_{2}(\mathbf{r})&=&0
\, .
\end{eqnarray}
\begin{figure}[!h]
\includegraphics[width=40mm,angle=-90]{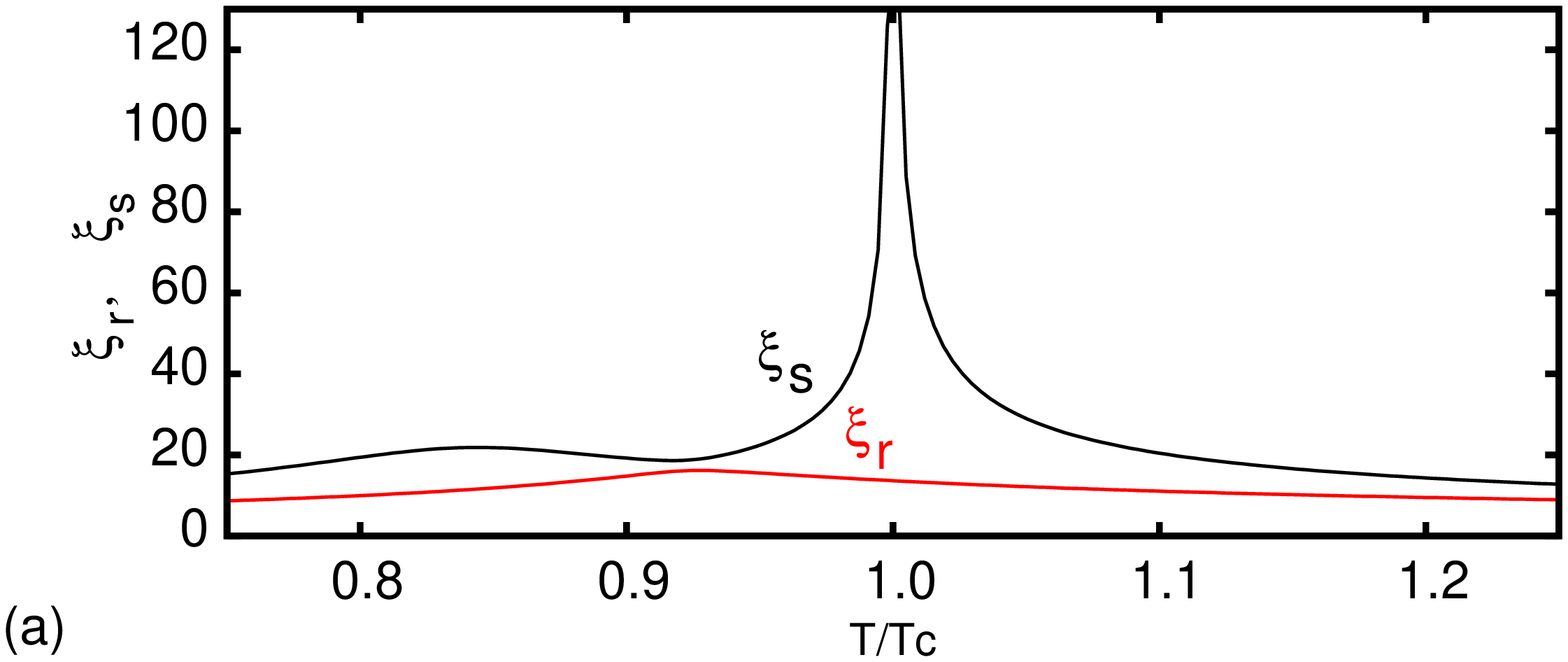}
\includegraphics[width=40mm,angle=-90]{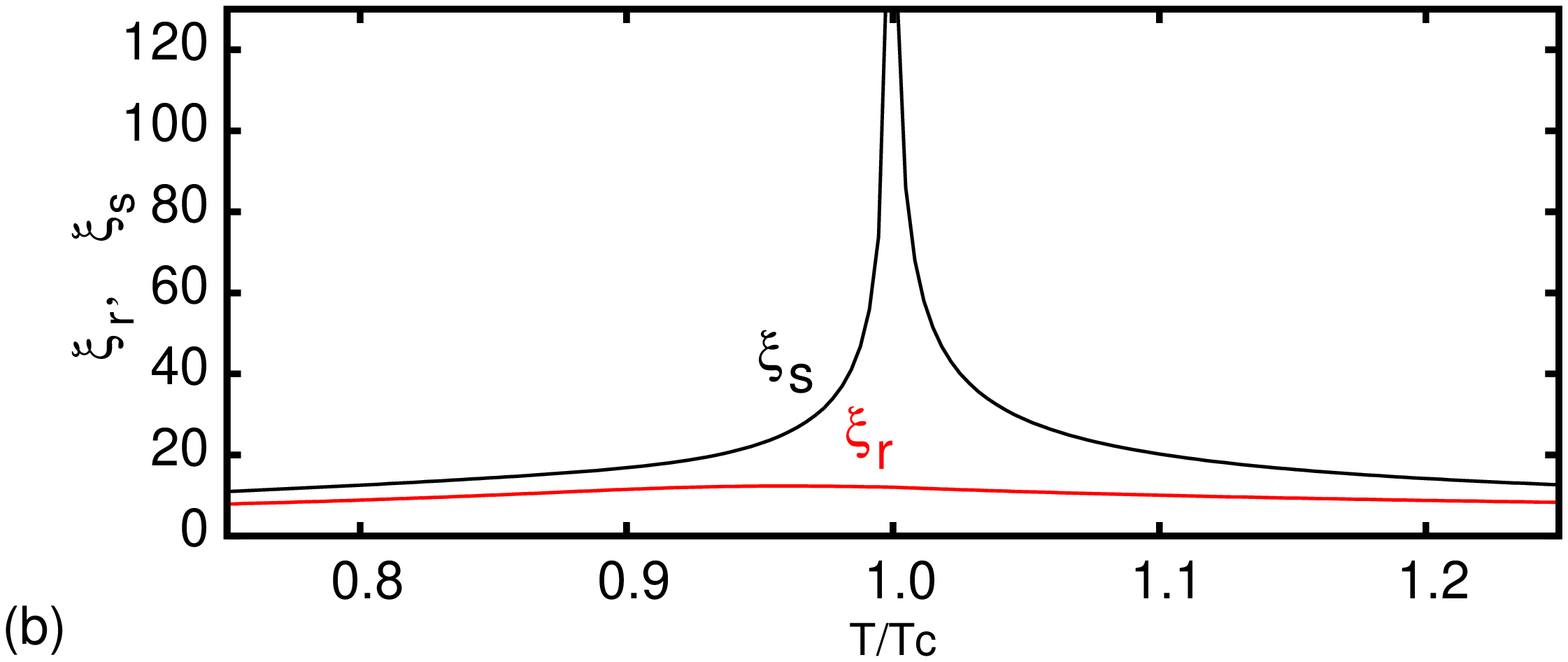}
\includegraphics[width=40mm,angle=-90]{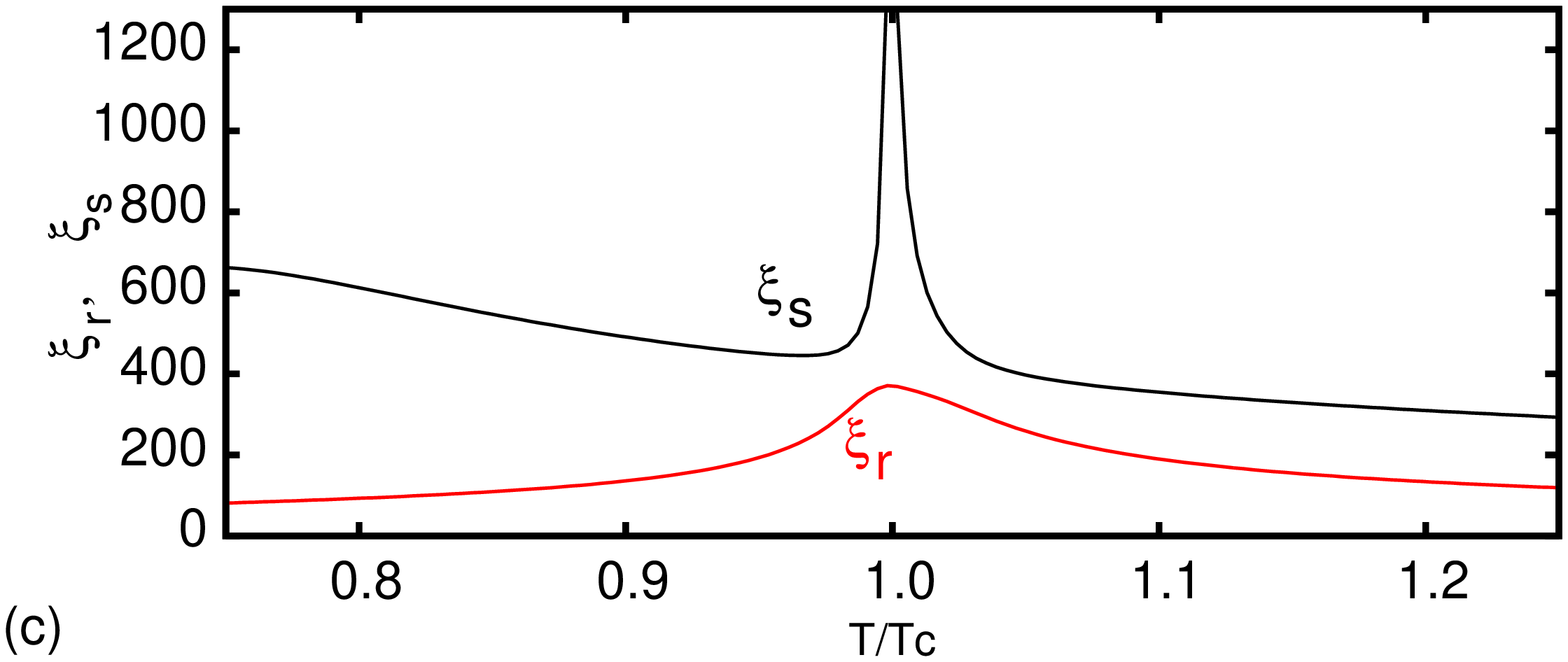}
\includegraphics[width=40mm,angle=-90]{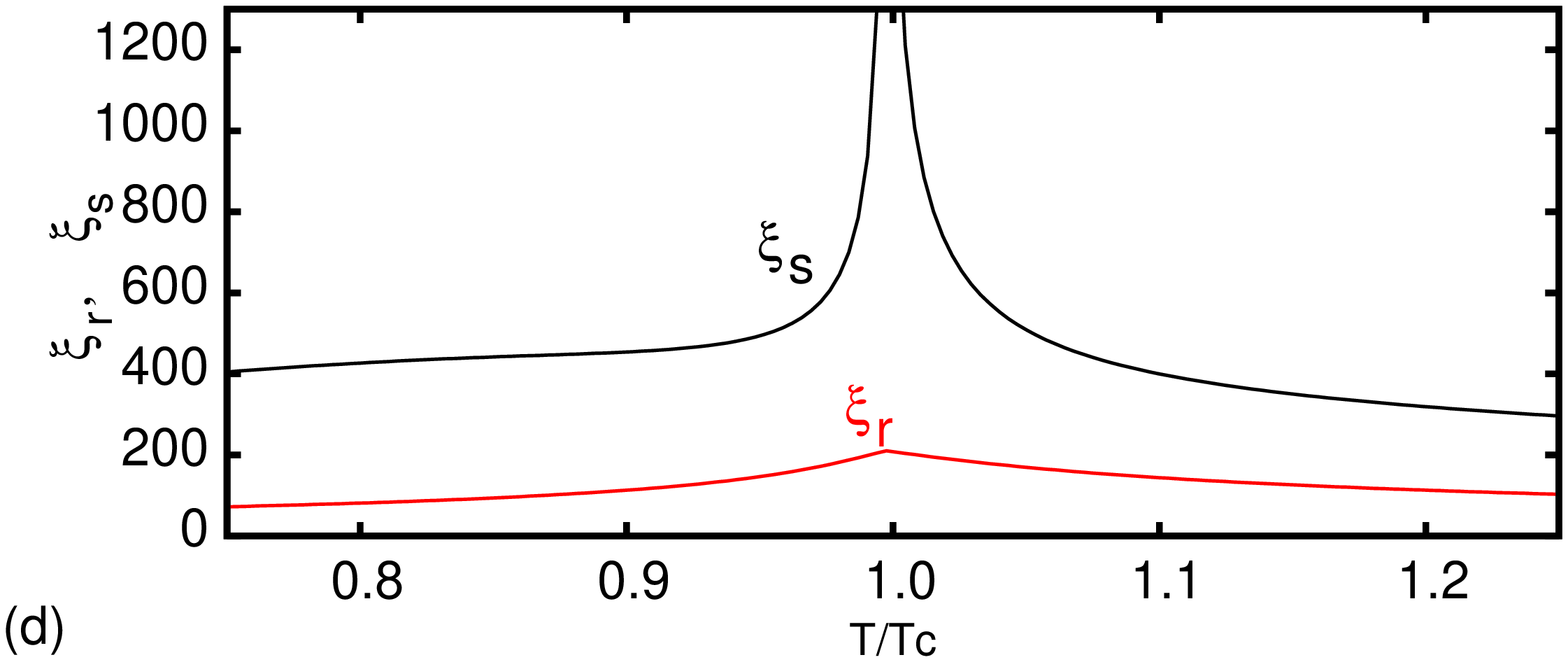}
 \caption{The soft and rigid coherence lengths ($\xi_r$, $\xi_s$)   versus temperature $T$ for several sets of parameters:
(a) and (c) $|U^{12}|=0.01$; (b) and (d) $|U^{12}|=0.04$; (a) and (b) $n=1.0$, (c) and (d) $n=0.75$.
$\xi_r$, $\xi_s$ are expressed in terms of lattice constant $a$.}
\end{figure}
the condition for the existence of non-trivial solutions of which,
\begin{eqnarray}\label{24}
&& \left|
\begin{array}{c}
  1+U^{11}\tilde{g}_{1}(T) +U^{11}\frac{\beta_{1}(T)}{\xi^{2}},~~ U^{12}\tilde{g}_{2}(T)
+U^{12}\frac{\beta_{2}(T)}{\xi^{2}} \\
  U^{21}\tilde{g}_{1}(T)
+U^{21}\frac{\beta_{1}(T)}{\xi^{2}},~~  1+U^{22}\tilde{g}_{2}(T)
+U^{22}\frac{\beta_{2}(T)}{\xi^{2}} \\
\end{array}%
\right|
\nonumber \\ && = 0
\, ,
\end{eqnarray}
determines the coherence lengths of the two-gap superconductor under consideration. On the basis of Eq. (\ref{24}) we have two solutions for $\xi$,
\begin{eqnarray}\label{25}
\xi_{s,r}^{2}\left(T\right)=\frac{G\left(T\right)
\pm\sqrt{G^{2}\left(T\right)-4K\left(T\right)\gamma(T)}}{2K\left(T\right)} \, ,
\end{eqnarray}
where
\begin{eqnarray}\label{26}
G\left(T\right)&=&\left(U^{12}\right)^{2}\left[\tilde{g}_{1}(T)\beta_{2}(T)+\tilde{g}_{2}(T)\beta_{1}(T)\right] \nonumber\\ &-&\left[1+U^{11}\tilde{g}_{1}(T)
\right]U^{22}\beta_{2}(T) \nonumber\\
&-&\left[1+U^{22}\tilde{g}_{2}(T)\right]U^{11}\beta_{1}(T) \, ,
\end{eqnarray}
\begin{eqnarray}\label{27}
K\left(T\right)&=&\left[1+U^{11}\tilde{g}_{1}(T)\right]\left[1+U^{22}\tilde{g}_{2}(T)\right] \nonumber\\
&-&\left(U^{12}\right)^{2}\tilde{g}_{1}(T)\tilde{g}_{2}(T) \, ,
\end{eqnarray}
\begin{eqnarray}\label{28}
\gamma(T)=\left[U^{11}U^{22}-\left(U^{12}\right)^{2}\right]\beta_{1}(T)\beta_{2}(T) \, .
\end{eqnarray}

The results of our specific two-dimensional lattice calculation of $\xi^s$ and $\xi^r$ versus $T/T_c$ are presented in Fig.
3.
 The soft or critical coherence length $\xi_{s}(T)$ diverges at the phase transition point $T=T_{c}$, while the rigid or non-critical coherence
length $\xi_{r}(T)$ remains finite (see Fig. 3). Furthermore, for fairly small inter-orbital interaction $|U^{12}|=0.01$ one can notice
that $\xi_s$ shows the second maximum (see the peak of $\xi_s$ in Fig. 3a and the rise of $\xi_s$ with lowering $T$ in Fig. 3c) as the memory
about the
lower transition
temperature in the intependent orbital (the one with
smaller intra-orbital
attraction). Such a behaviour is absent for stronger inter-orbital interaction.
Additionally, it is worth to notice that  the coherence length is much larger for $n=0.75$ that for $n=1.0$ (see Figs. 3a,b and 3c,d respectively).
This is  again the effect of the density
of states. Evidently, Van Hove singularity\cite{Ref markiewicz} scales strongly $T_c$ and other superconducting parameters including the coherence
length.

\section{Conclusions}

We considered two orbital superconductor with a small interaction coupling
producing possibility of the Copper pair transfer between orbitals. The system have been modelled by the negative-$U$ Hubbard model on the
two dimensional lattice. The obtained two values of the coherence lengths reflects the multi-orbital mechanism of superconductivity and
could have some consequence in specific spatial properties of such a complex supeconducting state \cite{babaev,annett}

Our results show that band filling strongly effects on the coherence lengths. Especially, Van Hove singularity present in the density of states plays an
important role in
scaling the coherence lengths.
The temperature dependence of the coherence lengths in our lattice model confirms the similar calculations in the continuous system\cite{ord1}.
One solution diverges near the phase transition point, while
the other one is non-critical.
The non-monotonic temperature dependence of coherence lengths is more pronounced if  the inter-orbital coupling is
sufficiently weak.

\begin{acknowledgements}
 This research was supported by the European Union through the European Regional Development Fund (Centre of Excellence "Mesosystems: Theory and
Applications", TK114). We
acknowledge the support by the Estonian Science Foundation, Grant No 7296. G.L. and T.\"{O}. kindly acknowledge a financial support by
the European Union,
under Grant Agreement No. UDA-POKL.04 01.01-00-108/08-00.
\end{acknowledgements}


\begin{thebibliography}{0}
\bibitem{spectrum}
All illustrative numerical calculations have been carried out for two-dimensional square lattice with hopping integrals between nearest neighbours $t^{\alpha\alpha}_{ij}=t$. The electron band energies associated with $s$-orbitals are $\varepsilon_{1}(\mathbf{k})=\varepsilon_{2}(\mathbf{k})=\varepsilon^{0}-2t\left[\cos (ak_{x}) + \cos (ak_{y})\right]$ with $-\frac{\pi}{a}\leq k_{x,y}\leq \frac{\pi}{a}$ where $a$ is the lattice constant. We have neglected the Hartree-type term in Eq. (\ref{4}).
\end{thebibliography}
\end{document}